\documentclass[twocolumn,times]{aastex62}

\usepackage{color}

\shorttitle{Galactic Halo Dispersion Measure}
\shortauthors{Yamasaki \& Totani}

\begin{document}

\title{THE GALACTIC HALO CONTRIBUTION TO THE DISPERSION MEASURE
  OF \\ EXTRAGALACTIC FAST RADIO BURSTS}

\correspondingauthor{Shotaro Yamasaki}
\email{yamasaki@astron.s.u-tokyo.ac.jp}

\author[0000-0002-1688-8708]{Shotaro Yamasaki}
\affiliation{Department of Astronomy, School of Science, The University of Tokyo,  7-3-1 Hongo, Bunkyo-ku, Tokyo 113-0033, Japan}

\author{Tomonori Totani}
\affiliation{Department of Astronomy, School of Science, The University of Tokyo,  7-3-1 Hongo, Bunkyo-ku, Tokyo 113-0033, Japan}
\affiliation{Research
Center for the Early Universe, School of Science, The University of
Tokyo, 7-3-1 Hongo, Bunkyo-ku, Tokyo 113-0033, Japan}

\begin{abstract}

  A new model of the Milky Way (MW) halo component of the dispersion
  measure (DM) for extragalactic sources, such as fast radio bursts
  (FRBs), is presented in light of recent diffuse X-ray observations.
  In addition to the spherical component of isothermal gas ($kT\sim0.3$
  keV) in hydrostatic equilibrium with the Galactic gravitational
  potential, our model includes a disk-like non-spherical hot gas
  component to reproduce the directional dependence of the observed
  X-ray emission measure (EM).  The total gas mass
  ($1.2\times10^{11}\,M_{\odot}$) is dominated by the spherical component, and
  is consistent with the total baryon mass of the MW expected from the
  dark matter mass and the cosmic baryon-to-dark-matter ratio. Our model predicts a mean halo DM of $43\:\,{\rm pc\:cm^{-3}}$, with a full range of $30$--$245\:\,{\rm
    pc\:cm^{-3}}$ over the whole sky.
  The large
  scatter seen in the X-ray EM data implies a $\sim0.2$ dex (rms)
  fluctuation of the MW halo DM.  We provide an analytic formula to estimate
  the MW halo DM of our model along any line of sight, which
  can be easily used to compute the total
  MW component of DM toward extragalactic sources, in combination with existing DM models of the warm
  ionized medium associated with the Galactic disk.
\end{abstract}


\defcitealias{Faerman2017}{F17}
\defcitealias{Fang2013}{F13}
\defcitealias{Nakashima2018}{N18}
\defcitealias{Cordes2002}{NE2001}
\defcitealias{Yao2017}{YMW16}
\defcitealias{Prochaska2019}{PZ19}

\section{Introduction}

The millisecond-duration radio transients, the so-called fast radio
bursts (FRBs) are one of the most enigmatic astronomical objects
(e.g., \citealt{Lorimer2007,Thornton2013}).  FRB observation is
a rapidly growing field; about 90 FRBs have been reported to date
\citep{Petroff2016}, but their origin and physical mechanism (see
\citealt{Platts2018,Cordes2019} for review) remain shrouded in a deep fog of
mystery. Interestingly, their dispersion measures ${\rm DM}\equiv \int
n_e ds$ (a line-of-sight integration of electron number density
$n_e$), typically hundreds of ${\rm pc \:cm^{-3}}$, are too large to
be attributed to free electrons in the Milky Way (MW), and a
cosmological distance scale of $z \sim$ 1 is inferred if the dominant
contribution to DMs is from electrons in the ionized intergalactic
medium (IGM). The cosmological origin has been first confirmed for a
repeating source FRB 121102 \citep{Spitler2016} with its localization
to a dwarf star-forming galaxy at redshift $z=0.19$
\citep{Chatterjee2017,Tendulkar2017}.
Recently, the host galaxies of two FRBs that have not yet repeated, 
FRB 180924 \citep{Bannister2019} and 190523 \citep{Ravi2019}, have
been identified to massive galaxies with no or weak star formation
at $z=0.32$ and $z=0.66$, respectively.
Even more recently, nine new repeating sources have been discovered by CHIME \citep{Amiri2018,Andersen2019} and repetitions have been confirmed for FRB 171019 originally singly-detected by ASKAP \citep{Kumar2019}, but their host galaxies remain unidentified.

In general, the observed total dispersion for an extragalactic source
${\rm DM}_{\rm obs}$ can be split into the four components as
\begin{eqnarray}
\label{eq: DM_host}
{\rm DM}_{\rm obs}={\rm DM}_{\rm ISM}+{\rm DM}_{\rm halo}+{\rm DM}_{\rm IGM}+{\rm DM}_{\rm host},
\end{eqnarray}
where ${\rm DM}_{\rm ISM}$ is the contribution from the warm ionized
medium (WIM; $T\lesssim10^4$ K) of interstellar medium (ISM) in the MW
disk, ${\rm DM}_{\rm halo}$ is that from the extended
hot Galactic halo ($T\sim10^6$--$10^7$ K), ${\rm DM}_{\rm IGM}$ is
that from IGM, and ${\rm DM}_{\rm host}$ is that from the host galaxy
including the local surrounding environment of the source.
Here we
neglected the contribution from intervening galaxy halos
\citep{McQuinn2014,Shull2018,Prochaska2018,Prochaska2019}.  In principle we can
estimate the contribution of MW electrons to total DM (i.e., ${\rm
  DM}_{\rm ISM}$ and ${\rm DM}_{\rm halo}$) for the direction of an
observed radio source, thereby obtaining a rough estimate of the
maximum source distance through the analytic ${\rm DM}_{\rm IGM}$--$z$
relation
(\citealt{Ioka2003,Inoue2004,Deng2014,McQuinn2014,Shull2018,Li2019},
for simulation study see also \citealt{Dolag2015,Pol2019}).  For the
ISM contribution, the warm electron density distribution models, such
as \citetalias{Cordes2002} \citep{Cordes2002,Cordes2003} and
\citetalias{Yao2017} \citep{Yao2017}, have been developed and widely
used, whereas less attention has been paid to the smaller halo
contribution (often ignored for the sake of simplicity indeed).

The distribution of hot gas ($k T \sim 0.3$ keV) in the MW halo has
been studied based on analytic gas density profile or numerical
simulations, with observational constraints from oxygen absorption
lines in UV or X-ray bands, emission measure (EM) of diffuse X-ray
emission, and DM toward the Large Magellanic Cloud (LMC)
(\citealt{Maller2004,Sommer2006,Yao2009,Fang2013,Nuza2014,Dolag2015,Tepper2015,Roca2016,Faerman2017,Fielding2017,Li2017,Nakashima2018,Shull2018,Prochaska2019}).
The DM value estimated by these modelings is ${\rm DM}_{\rm halo}\sim
30$--$80\: {\rm pc\:cm^{-3}}$.  Most of these studies considered a
spherically symmetric halo, but recent X-ray observations of diffuse
halo gas revealed a significant directional dependence of the EM, which
motivated several studies to introduce a disk-like halo gas
distribution (\citealt{Yao2009,Fang2013,Li2017,Nakashima2018}). It should be noted that this hot disk-like halo component is completely different in physical properties (such as temperature and geometrical shape) from the so-called ``warm thick disk'', which is included in the warm ISM models and generally constrained by Galactic pulsar DM measurements (see Section \ref{subsec:Discussion ISM}).

However, such a disk-like model results in a scale
radius of less than 10 kpc and the associated gas mass much smaller than that of total
halo gas expected from the total dark matter mass of the MW halo and the
cosmic ratio of baryons to dark matter.  This indicates that we need
to incorporate two components for a realistic model of MW halo gas
distribution: a spherical component extending up to the virial radius
($\sim$200 kpc) and a more compact disk-like component responsible for
the diffuse X-ray emission\footnote{ \citet{Fang2013} explored the Galactic gas distribution by combining warm thick disk and hot spherical halo, which is in contrast to our idea of combining two hot gas halo components.}. Both components may have significant
contribution to the DM, and the purpose of this work is to construct
such a two-component, direction-dependent model
of the MW halo gas distribution and DM.

This paper is organized as follows.  In Section \ref{sec:models}, we
describe the framework of our model for spatial distribution of the hot gas halo. The modeled EM and DM are compared with
observational constraints in Section \ref{sec:constraints}. We then
provide a fitting formula of the halo DM as a function of the Galactic
coordinate for a convenient use in FRB observations in Section
\ref{sec:Analytic formula}, and discussion on our newly proposed model
is given in Section \ref{sec:Discussion}, followed by conclusions in
section \ref{sec:Conclusions}. The adopted cosmological parameters for
a flat universe are $H_0 = 67.8\,{\rm km\, s^{-1}\,Mpc^{-1}}$,
$\Omega_{m} = 0.308$, $\Omega_{\Lambda}= 0.692$ and $\Omega_{b}=
0.0483$ \citep{Planck2016}. Our choice of the Galactocentric distance
of the Sun is $D_{\odot}=8.5$ kpc \citep{Kerr1986}.
When calculating the
number density of gas particles, we take a mean molecular mass per
electron, $\mu_e\equiv\rho/(m_pn_e)=1.18$ ($\rho$, $m_p$ and $n_e$ are
gas mass density, the proton mass and electron number density,
respectively), a mean particle mass $\mu\equiv\rho/(m_pn)=0.62$ ($n$
denotes the number density of all particles, including baryonic
particles and free electrons, that contribute to the gas pressure),
and a number density ratio of hydrogen to electron
$\chi_{\rm{H}}=0.82$ (independent of gas metallicity).
These were calculated assuming fully 
ionized hydrogen and helium with a helium mass
abundance of 30\%.

\begin{figure*}[t]
\epsscale{1.1}
\plotone{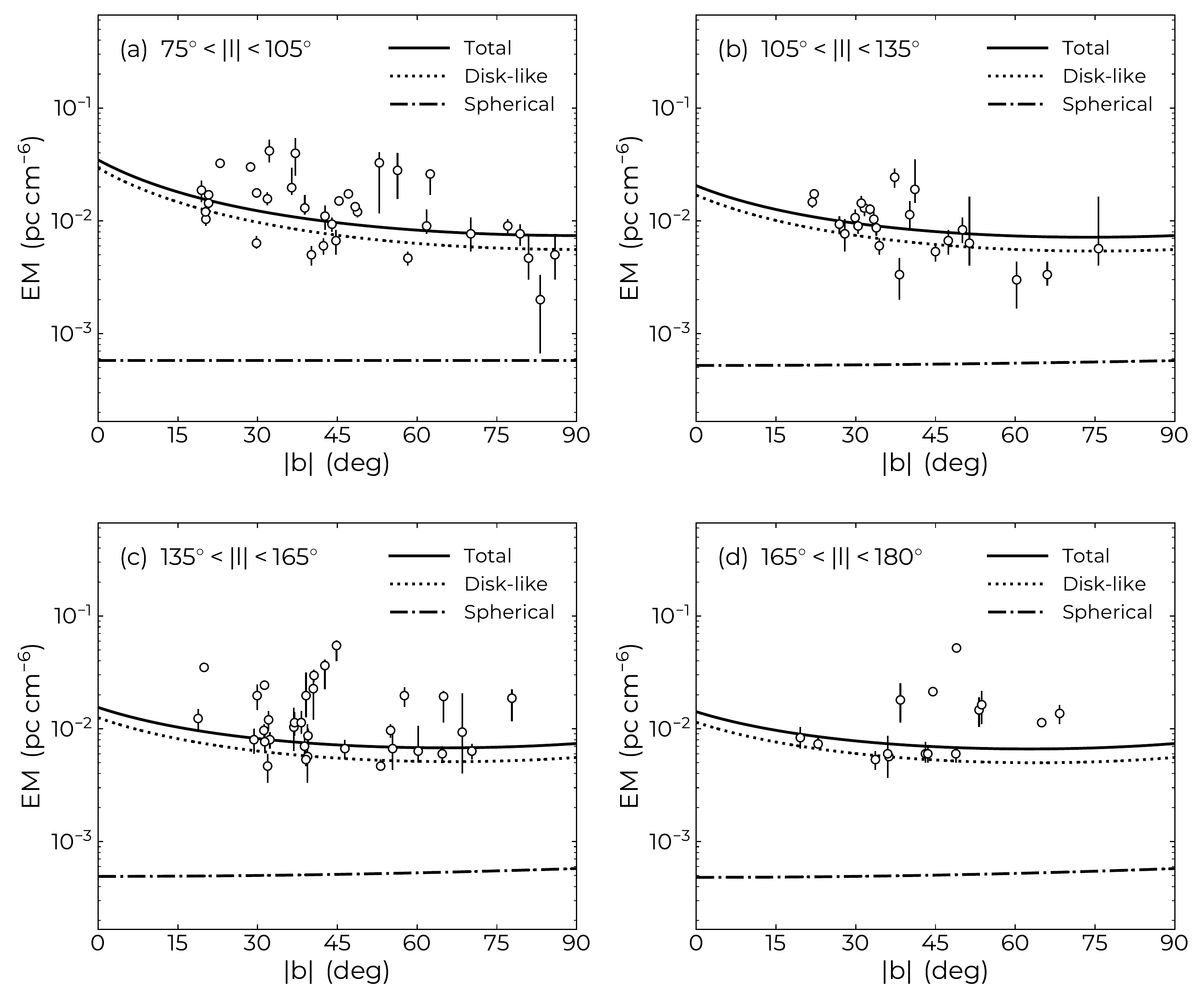}
\caption{Emission measures of the hot halo gas as a function of
  Galactic latitude. Each panel corresponds to four different regions
  in Galactic longitude. The halo gas metallicity is assumed to be $Z_{\rm halo}
  = 0.3 Z_\odot$ for the observed EM data points.  Two data points in
  107 sightlines in Table 1 of \citet{Nakashima2018} with an upper limit are removed in this figure 
  and model fitting. Model predictions are plotted for (a)
  $|l|=90^{\circ}$; (b) $|l|=120^{\circ}$; (c) $|l|=150^{\circ}$; (d)
  $|l|=180^{\circ}$. Here $|l|$ is defined such that $|l|= l\:\,
  (0^{\circ}\le l \le 180^{\circ})$ and $|l|= 360^{\circ}-l$
  (otherwise).  }
\label{fig:EM_vs_b}
\end{figure*}

\section{A Model for the hot gas halo }
\label{sec:models}

Hot gas existing in the MW halo can be probed by EM (${\rm EM} \equiv
\int n_e n_{\rm H} ds$) of diffuse X-ray emission, where $n_{\rm H}$
is the hydrogen number density and $s$ is a coordinate along the line
of sight (e.g.,
\citealt{Snowden1997,McCammon2002,Yao2007,Yao2009,Gupta2012,Yoshino2009,Hagihara2010,Henley2013,Nakashima2018}).
Most recently, \citet{Nakashima2018} (hereafter
\citetalias{Nakashima2018}) have estimated EM to 107 sightlines by the
           {\it Suzaku} X-ray observations at
           $75^{\circ}<l<285^{\circ}$ and $|b|>15^{\circ}$.  They
           found that the observed EM distribution over the entire sky
           cannot be descirbed by a spherically symmetric electron
           density distribution, but can be reproduced by a disk-like
           distribution
\begin{equation}
\label{eq:N18_disk}
n_e^{\rm disk}(R,z)=n_0^{\rm disk}
\exp{\left[-\left(\frac{R}{R_0}+\frac{|z|}{z_0}\right)\right]},
\end{equation}
where $R$ and $z$ are the cylindrical coordinates, and $n_0^{\rm disk}=3.8_{-1.2}^{+2.2}\times10^{-3}
(Z_{\rm halo}/Z_\odot)^{-1} \ {\rm cm^{-3}}$ ($Z_{\rm halo}$ denotes the halo hot gas
metallicity), $R_0=7.0_{-1.7}^{+2.1}\:{\rm kpc}$,
and $z_0=2.7_{-0.7}^{+0.8}\:{\rm kpc}$. The metallicity dependence
appears because X-ray emissivity is dominated by oxygen ions.

However, the total mass of this disk-like component is only $\sim 2
\times 10^8 (Z_{\rm halo}/Z_\odot)^{-1} M_\odot$, which is much
smaller than the total halo gas mass expected by the MW dark halo mass
and the cosmic mass ratio of dark to baryonic matter (e.g.,
\citealt{Yao2007,Fang2013,Miller2015,Li2017}). Although such a more
massive, more spherical, and more extended (up to the virial radius)
halo may not significantly contribute to the observed EM of diffuse
X-ray emission, it should exist theoretically (e.g.,
\citealt{Spitzer1956,Cen1999}) and it is also supported by
observations of absorption lines (e.g.,
\citealt{Nicastro2002,Tumlinson2011}).  Therefore in this work we
perform a new fit to the observed EM of \citetalias{Nakashima2018} with the two components of
the compact disk-like halo and the extended spherical halo.

\begin{figure*}[t]
\epsscale{1.1}
\plotone{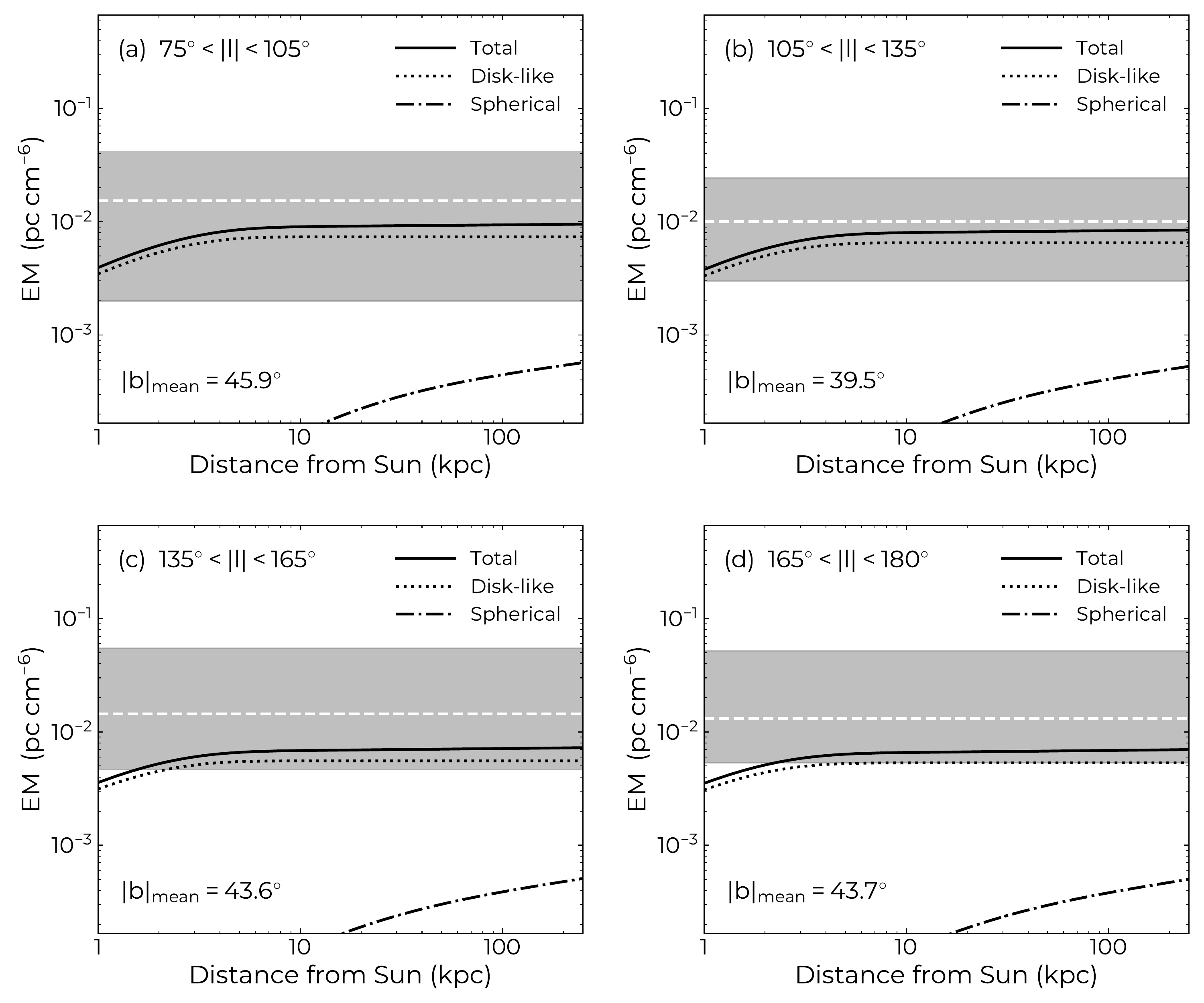}
\caption{Emission measures of the hot gas halo as a function of
  radius.  The grey shaded region denotes the full range of variation
  in data, and the horizontal dashed line is the mean.  Model
  predictions are plotted for the same Galactic longitude as Figure
  \ref{fig:EM_vs_b} with Galactic latitude chosen to the mean of data
  $|b|_{\rm mean}$ as shown in each panel.}
\label{fig:EM_profile}
\end{figure*}

For the spherical component, we therefore introduce a theoretical
density profile that is modeled as the isothermal gas in hydrostatic
equilibrium with a Galactic dark matter halo.  In our model, MW's dark
matter halo has a virial mass of $M_{\rm vir}=10^{12}\,M_{\odot}$ and
a virial radius of $r_{\rm vir}=260$ kpc according to the
model by \citet{Klypin2002}. We assume that the dark matter
distribution follows the NFW-profile gravitational
potential $\Phi$ with a concentration
of $c_{\rm vir}\equiv r_{\rm vir}/r_s=12$ with $r_s$ being a NFW scale
radius (\citealt{NFW1997,Bullock2001}). The gravitational potential of
the Galactic stellar disk is neglected since it has little effect on
the resulting density profile. Assuming the ideal gas with a constant
temperature $T_{\rm halo}$, gas pressure is given by
$P/\rho=kT_{\rm halo}/(\mu m_p)$
and hydrostatic equilibrium (HE) $\nabla
P = -\rho \nabla \Phi$ reduces to
\begin{eqnarray}
\label{eq:isothermal_sphe}
n_e^{\rm sphe}(r) &=& n_0^{\rm sphe} 
\exp{\left\{-\Upsilon\left[1-\frac{\ln{(1+r/r_s)}}{r/r_s}
    \right]\right\}},
\end{eqnarray}
where $\Upsilon=4\pi G r_s^2\rho_s\mu m_p/(kT_{\rm halo})$ is a
dimensionless constant with $\rho_s = \rho(r_s)$ being the NFW scale
density. We assume the same temperature for the disk-like and
spherical components as $kT_{\rm halo}=0.3$ keV based on the X-ray
observation \citepalias{Nakashima2018}.  The central electron density
$n_0^{\rm sphe}$ is determined so that the enclosed gas mass of the
spherical component within $r_{\rm vir}$ is equal to the Galactic
baryon mass $M_b$:
\begin{eqnarray}
\label{eq:total gas mass}
\int_0^{r_{\rm vir}} 4\pi 
r^2 \mu_e m_p \,n_{e}^{\rm sphe}(r)\,dr=M_b.
\end{eqnarray}
We choose the fiducial total baryon mass to be $M_b =1.2\times
10^{11}\,M_{\odot}$, assuming the baryon fraction $M_b/M_{\rm vir}$ of the
MW is $\sim75$\% of the cosmic mean $\Omega_b/\Omega_m\sim0.16$, which
is the same value as adopted by
\citet{Prochaska2019}. These figures are roughly consistent with an
estimate that $\sim26$\% of
galactic baryons reside in the stars and ISM \citep{Fukugita1998}, if the remaining $\sim74$\% of baryons are in the galactic halo. Combining the above assumptions, we obtain $\Upsilon=2.6$ and $n_0^{\rm sphe}=3.7\times10^{-4}\: {\rm cm^{-3}}$ as fiducial values for our model. 

\begin{figure*}[t]
\epsscale{1.15}
\plotone{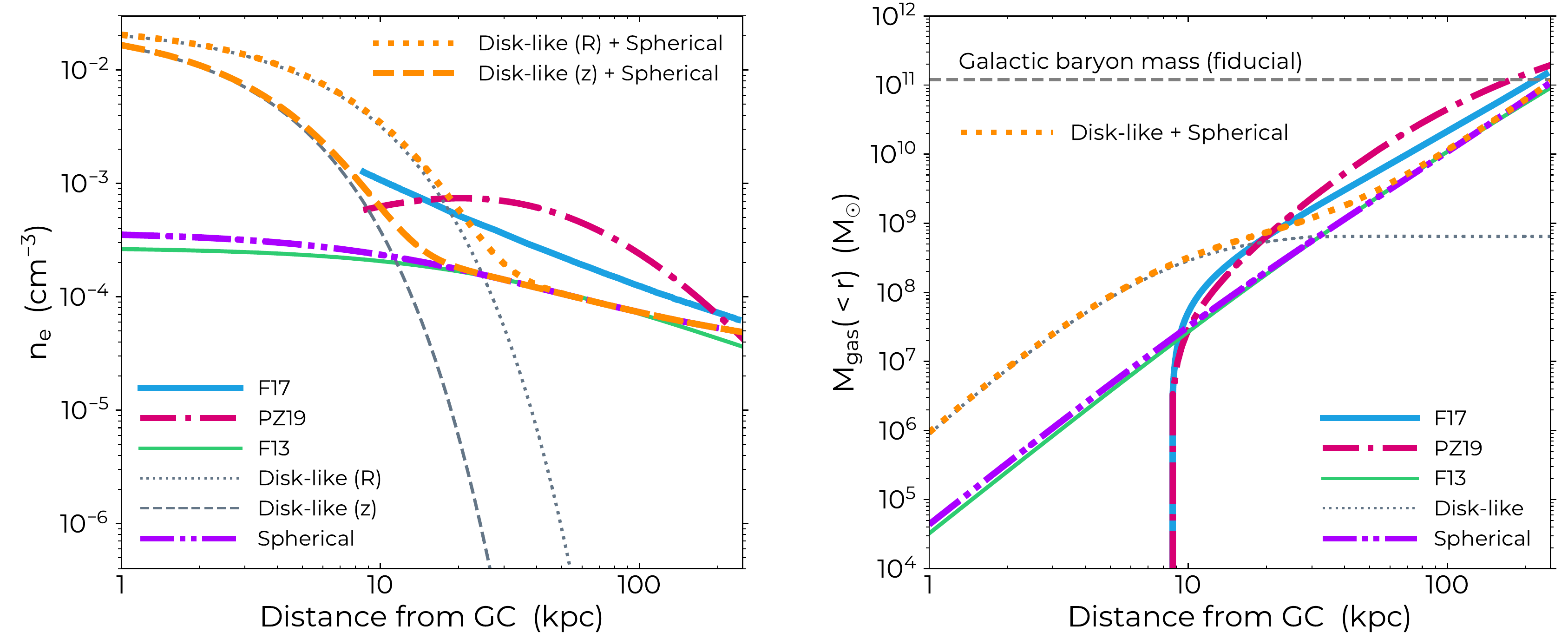}
\caption{Hot gas profile as a function of Galactocentric radius. {\it
    Left panel}: electron density; {\it right panel}: enclosed hot gas
  mass within Galactocentric radius $< r$. For density profiles of the
  disk-like halo component of our model, two profiles into vertical
  ($z$-axis) and in-plane ($R$-axis) directions are shown.  The
  density profiles of \citetalias{Faerman2017} and
  \citetalias{Prochaska2019} are shown only at $r>D_{\odot}$,
  according to their definitions.
  The horizontal grey dashed line
  shown in the right panel indicates the fiducial baryon mass of the
  MW.}
\label{fig:ne_mass}
\end{figure*}

\section{Observational constraints}
\label{sec:constraints}
Here we construct our non-spherical hot gas halo models by fitting
to the observed X-ray EM.  Then we present our model in comparison with
existing theoretical models, and examine its consistency with DM of
LMC pulsars and absorption line observations.

\subsection{Fit to X-ray EM}
\label{subsec: X-ray observations}

Observations of the diffuse X-ray EM have benefits of a large number
of sightlines.  To construct a direction-dependent model of the MW
halo DM, we utilize the measurements of halo gas EM, EM$_{\rm N18,
  \odot}$, which are presented as EM$_{\rm halo}$ in Table 1 of \citetalias{Nakashima2018}.  The EM$_{\rm N18, \odot}$ for each siteline
was determined, along with gas temperature and [O/Fe], by spectral
fittings.
The medians of temperature and [O/Fe] over all
sitelines are $0.26$ keV and $0.25$, respectively
\citepalias{Nakashima2018}.
The metallicity of the halo gas is not well constrained by
the X-ray data, and it was fixed to the solar abundance.
However, X-ray emissions are dominated by continuum
recombination emission from oxygen ions. Therefore
the true EM should scale with the halo gas metallicity as:
\begin{eqnarray}
  {\rm EM}_{\rm N18} = \left(\frac{Z_{\rm halo}}{Z_\odot}\right)^{-1}
  {\rm EM}_{\rm N18, \odot} \ .
\end{eqnarray}
Throughout this work, we adopt $Z_{\rm halo} = 0.3\, Z_\odot$ as
suggested by cosmological simulations \citep{Cen2006} and observations
of high velocity clouds (e.g.,
\citealt{Gibson2000,Fox2005}). Distribution of EM$_{\rm N18}$ is shown
in four panels of Figure \ref{fig:EM_vs_b} as a function of Galactic
latitude.  Despite the large scatter seen in the data,
\citetalias{Nakashima2018} statistically confirm the trend of
decreasing EMs as Galactic latitudes increase.

We construct an empirical model for the entire electron density
distribution of the hot gas by combining Equation (\ref{eq:N18_disk})
and (\ref{eq:isothermal_sphe}): $n_e=n_e^{\rm disk}+n_e^{\rm
  sphe}$. Since the mass ratio between the disk-like and spherical
halo components within $r_{\rm vir}$ is expected to be small, a naive
summation of these barely affects the HE assumption.  The modeled EM
of the halo gas towards a given Galactic coordinate $(l,b)$ is
computed by
\begin{eqnarray}
\label{eq:EM_MW(halo)}
{\rm EM}_{\rm model}(l,b)&\equiv&\int_0^{s_{\rm max}} n_{e}(s)\,n_{\rm H}(s)\,ds ,
\end{eqnarray}
where $n_{\rm H}=\chi_{\rm H}\,n_e$ is the hydrogen number density and
we integrate the hot gas halo density profile along the line-of-sight
from the solar system 
out to the maximum distance $s_{\rm max}(l,b)$ corresponding to
the virial radius of the MW halo.

Since the spherical component is already fixed by Equation
(\ref{eq:total gas mass}), the remaining parameters to be determined
are $n_0^{\rm disk}$, $R_0$ and $z_0$ that characterize the disk-halo
component. These are determined by fitting EM$_{\rm model}$
to EM$_{\rm N18}$ using Markov Chain Monte
Carlo (MCMC) sampling with {\fontfamily{qcr}\selectfont emcee}, a
Python based affine invariant sampler \citep{Foreman-Mackey2013}.  The
likelihood function is $\ln\left({\cal L}\right)=-\chi^2/2$, where the
standard deviation associated with each data is defined by the
geometric mean of the asymmetric errors.  We adopt a flat prior
distribution for all of our parameters in the $n_0^{\rm
  disk}/(10^{-3}\:{\rm cm^{-3}})\in[0.1,100]$, $R_0/({\rm
  kpc})\in[0.1,100]$, and $z_0/({\rm kpc})\in[0.1,100]$.  We generate
$10^5$ samples and obtain the best set of parameters with
$n_0^{\rm disk}=7.4_{-1.6}^{+2.2}\times10^{-3} (Z_{\rm halo}/Z_\odot)^{-1}
\:{\rm cm^{-3}}$,
$R_0=4.9_{-0.5}^{+0.6}\:{\rm kpc}$, $z_0=2.4_{-0.4}^{+0.4}\:{\rm kpc}$, where errors are estimated by the $16$th and $84$th percentile of the MCMC realizations.

The EMs predicted with best-fit parameters as a function of Galactic
latitude and distance are presented in Figure \ref{fig:EM_vs_b} and
Figure \ref{fig:EM_profile}, respectively.  In Figure
\ref{fig:EM_vs_b} we see that the
modeled EM is dominated by the disk-like component, and 
our model matches the trend of EM against spatial directions.
Figure
\ref{fig:EM_profile} indicates that the total EM reaches the observed
EM at $\lesssim5$ kpc from the Sun, reflecting the dominance of
the disk-like halo.  The large scatter of the data from the mean might
be due to a density fluctuation in the disk-like halo.

\subsection{Comparison with Previous Studies}
\label{subsec: Comparison with Previous Studies}

Here we compare our model with three models of previous studies in the
literature: (1) an isothermal model with multiple gas phases
(the fiducial model with $Z_{\rm halo}=0.5\,Z_{\odot}$ of
\citealt{Faerman2017}; hereafter \citetalias{Faerman2017}), (2) an
adiabatic gas model with polytropic index $5/3$ (\citealt{Fang2013},
hereafter \citetalias{Fang2013}, see also \citealt{Maller2004}) and (3) a modified NFW
profile with $\alpha=y_0=2$ of
\citet{Mathews2017} and \citet{Prochaska2019} (hereafter
\citetalias{Prochaska2019}).
Figure \ref{fig:ne_mass} shows density (left panel) and mass (right
panel) profile of the hot gas for different models. Since the
disk-like component in our model has directional dependence, profiles
to two directions (along $R$- and $z$-axis) are shown.  The spherical
component of our model is quantitatively similar to the \citetalias{Fang2013} model.

\begin{figure}[t]
\epsscale{1.15}
\plotone{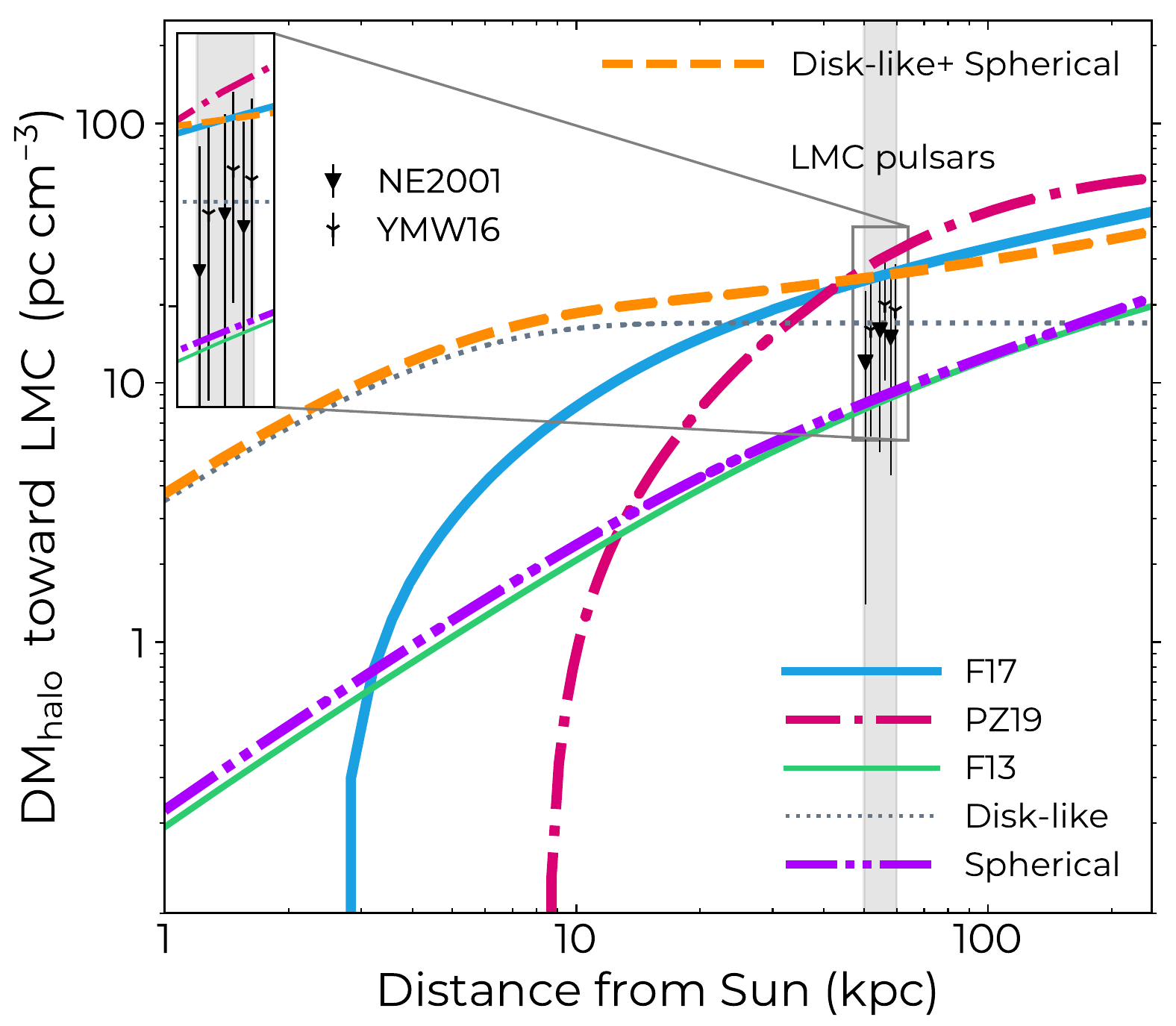}
\caption{Hot gas halo dispersion measure as a function of distance in
  the LMC direction. LMC pulsar data are randomly distributed at $s\in
  [50,60]$ kpc (denoted by the grey shaded region)
  for display purposes. The two data points for the same pulsar
  but assuming the two different DM models of the Galactic disk
  are placed next to each other. The region around the data points
  is also shown as inset zoom-in.
}
\label{fig:DM_profile}
\end{figure}

\subsection{LMC Pulsar Dispersion Measure}
\label{subsec: Pulsar dispersion measures}

The dispersion measurements of pulsars outside the Galactic disk
provide us the most direct tool to reveal the gas distribution in the
MW. While most Galactic pulsars lie in the Galactic disk, some of them
have been found in the Large Magellanic Cloud (LMC) at a distance of
$49.97\pm 1.3$ kpc \citep{Pietrzynski2013} from the Sun. Because of
the large offsets from the Galactic disk, LMC pulsars have been used
to probe the hot halo gas distribution in the literature (e.g.,
\citealt{Anderson2010}; \citetalias{Fang2013}).  Here we focus on three LMC pulsars
with the lowest DMs of $65,\,68,\,69\:\,{\rm pc\:cm^{-3}}$
\citep{McConnell1991,Manchester2006}. To estimate the hot gas halo
contribution to LMC pulsar DMs, we need to subtract the contribution
from the ISM in the MW disk and spiral arms. We find ${\rm DM_{\rm
    ISM}}=53 \:\,{\rm pc\:cm^{-3}}$ for \citetalias{Cordes2002} model
($48 \:\,{\rm pc\:cm^{-3}}$ for \citetalias{Yao2017} model), thereby
obtaining upper-limits on the hot gas halo DM at the LMC location as
${\rm DM_{\rm halo}}={\rm DM_{\rm PSR}}-{\rm DM_{\rm
    ISM}}=12$--$16\:\,{\rm pc\:cm^{-3}}$ for \citetalias{Cordes2002}
($17$--$21\:\,{\rm pc\:cm^{-3}}$ for \citetalias{Yao2017}). Here we
assume that the measurement uncertainty and the DM contribution from
local gas within the LMC are both negligible. The uncertainty arising
from ${\rm DM_{\rm ISM}}$ models is conservatively taken to be $20$\%
\citep{Cordes2002}.

Figure \ref{fig:DM_profile} shows the hot gas halo DM profiles along
the LMC sightline ($l=280^{\circ},\,b=-32.9^{\circ}$) for different
halo gas models.  Given the large uncertainty of warm ISM models, most
of the models are marginally consistent with the upper-limits
established by the LMC pulsar DM. It should be noted that 
the DM predicted by our model is dominated by
the disk-like component, which is not taken into account
in previous models. This implies that
gas density of the previous models by 
\citetalias{Faerman2017} and \citetalias{Prochaska2019} is 
too high as that of the spherical component of our model.

\subsection{Absorption Line Measurements}
\label{subsec: absorption lines}

Another observational constraint comes from the X-ray absorption lines
of highly ionized oxygen (O {\sc vii} and O {\sc viii}) and a UV
absorption line (O {\sc vi}) by MW halo gas seen in distant active
galactic nuclei (AGN) or blazar spectra (e.g.,
\citealt{Nicastro2002,Fang2002,Fang2003,Rasmussen2003,Sembach2003,Collins2004,McKernan2004,Wang2005,Bregman2007,BL2007,Hagihara2010,Gupta2012,Miller2013,Miller2015,Fang2015}). These observations have been used to constrain the
MW halo gas models (e.g., \citetalias{Faerman2017} and \citetalias{Prochaska2019}).

However, we only utilize data of the diffuse X-ray EM in this study
for the following reasons.  First, since the absorption lines
are not fully resolved with grating spectrometers, the same line
velocity width often needs to be assumed for different sitelines, and
thus the inferred ionic column density (i.e., DM) strongly depends on
the assumed gas kinematics.
The other reason is that absorption line measurements tend to be
limited to the direction of bright AGNs or blazars, which would lead
to a smaller size of sample per each measurement ($N\lesssim30$,
\citealt{Gupta2012,Miller2013,Miller2015,Fang2015}) compared to
diffuse X-ray observations ($N\gtrsim100$,
\citealt{Henley2013}; \citetalias{Nakashima2018}).
Therefore we chose to fix the spherical component of our
model by the total gas mass theoretically expected from
the MW dark mass. 

\newcommand\T{\rule{0pt}{2.6ex}}       
\newcommand\B{\rule[-1.2ex]{0pt}{0pt}} 

\begin{table*}
\smallskip 
\begin{center}
\caption{Coefficients $c_{ij}$ (in units of ${\rm pc\:cm^{-3}}$)
    in Equation (\ref{eq:polynomial}).}
\label{tab:coefficients}
\begin{tabular}{ccccccccc}
\hline\hline
\multicolumn{9}{c}{Disk-like $+$ Spherical  ($n=7$)} \T\B \\
\cline{2-9}   
$c_{ij}$ & $j=0$ & $j=1$ & $j=2$ & $j=3$ & $j=4$ & $j=5$ & $j=6$ & $j=7$\T\B \\ 
 \hline \T
$i=0$    & $250.12$ & $-871.06$ & $1877.5$ & $-2553.0$ & $2181.3$ &  $-1127.5$  &  $321.72$ & $-38.905$\\
$i=1$    & $-154.82$ &  $783.43$  &  $-1593.9$ &  $1727.6$ & $-1046.5$ &  $332.09$ & $-42.815$  & $0$  \\
$i=2$    & $-116.72$ &  $-76.815$  &  $428.49$ &  $-419.00$ &  $174.60$ & $-27.610$ & $0$  & $0$\\ 
$i=3$    & $216.67$ &  $-193.30$  &  $12.234$ &  $32.145$ & $-8.3602$ & $0$ & $0$ & $0$ \\
$i=4$    & $-129.95$ &  $103.80$  &  $-22.800$ & $0.44171$ & $0$ & $0$ & $0$  & $0$  \\
$i=5$    & $39.652$ &  $-21.398$ &  $2.7694$  & $0$ & $0$ & $0$ & $0$ & $0$   \\
$i=6$    & $-6.1926$ & $1.6162$  &  $0$  & $0$ & $0$ & $0$ & $0$ & $0$\B\\
$i=7$    & $0.39346$ &  $0$ &  $0$  & $0$ & $0$ & $0$ & $0$ & $0$\B\\
\hline
\multicolumn{9}{c}{Spherical   ($n=3$)}\T\B \\
\cline{2-9}
$c_{ij}$ & $j=0$ & $j=1$ & $j=2$ & $j=3$ & -- & -- & --& -- \T\B \\
\hline \T
$i=0$    & $25.325$ & $-1.4255$ & $-1.0546$ & $0.17295$ &--  &  -- &--& --   \\
$i=1$   &$-2.1749$ &  $1.4311$  &  $0.44722$ &  $0$ & -- & -- & -- & -- \\
$i=2$    & $-0.37683$ &  $-0.27977$  &  $0$ &  $0$ &-- & -- &-- & --  \\ 
$i=3$    & $0.16103$ &  $0$  &  $0$ &  $0$ &-- &-- &--& -- \B  \\
\hline
\end{tabular}
\end{center}
\end{table*}

One of the most recent study of OVII absorptions toward nearby AGNs at high Galactic latitudes $|b|\gtrsim30^{\circ}$ (\citealt{Fang2015}) suggests that typical column densities through the Galactic halo are $N_{\rm OVII} = 10^{15.5}$--$10^{17} \:{\rm cm^{-2}}$ with large scatters likely due to the measurement uncertainties. Assuming that O {\sc vii} is the dominant state among ionized oxygens with the hot gas metallicity of $Z_{\rm halo} = 0.3Z_{\odot}$, this translates into a dispersion measure of
\begin{equation}
{\rm DM}_{\rm OVII} = 82\:{\rm pc\, cm^{-3}} \left(\frac{N_{\rm OVII}}{10^{16.5}\:{\rm cm^{-2}}}\right) \left(\frac{Z_{\rm halo}}{0.3 Z_{\odot}}\right)^{-1},
\end{equation}
where we adopt a rough median value for $N_{\rm OVII}$ (see also \citealt{Shull2018} and \citetalias{Prochaska2019} for similar estimates). Meanwhile, our model predicts ${\rm DM}_{\rm halo} = 30$--$70 {\:\rm pc\, cm^{-3}}$ over the same region of the sky. 
Therefore, the difference between \citetalias{Faerman2017}, \citetalias{Prochaska2019}, and
the spherical component of our model is within the uncertainty
in using absorption lines, and hence our model is fully consistent
with the absorption line observations.

\begin{figure}[]
\epsscale{1.15}
\plotone{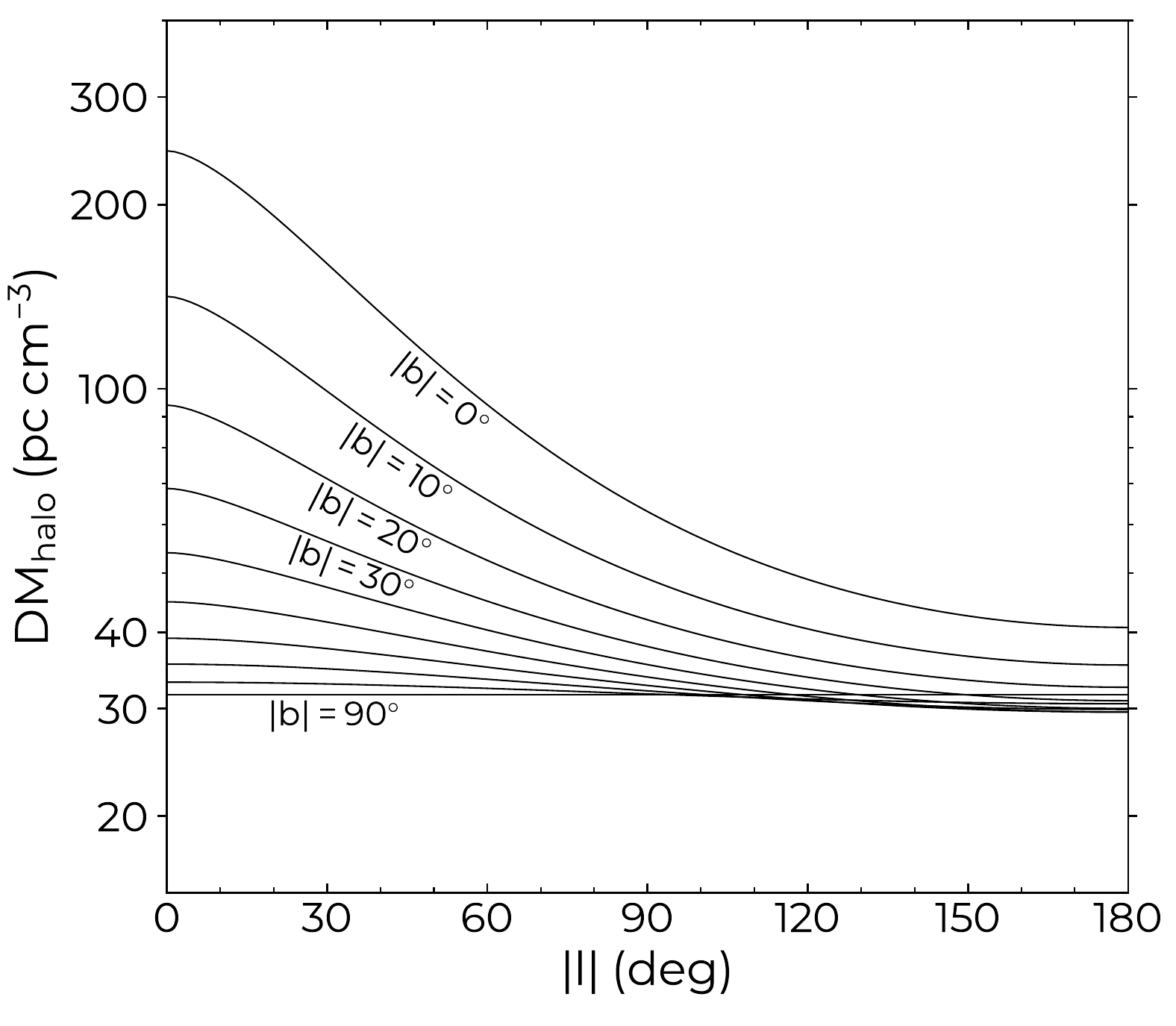}
\caption{Hot gas halo DM (disk-like halo plus spherical) as a function
  of the Galactic longitude. Ten curves are shown 
  corresponding to the Galactic latitude of 
  $|b| = 0^{\circ}$ to $90^{\circ}$ with a step of $10^{\circ}$.}
\label{fig:DM_halo}
\end{figure}

\section{Analytic formula for MW halo DM}
\label{sec:Analytic formula}

\begin{figure}[]
\epsscale{1.15}
\plotone{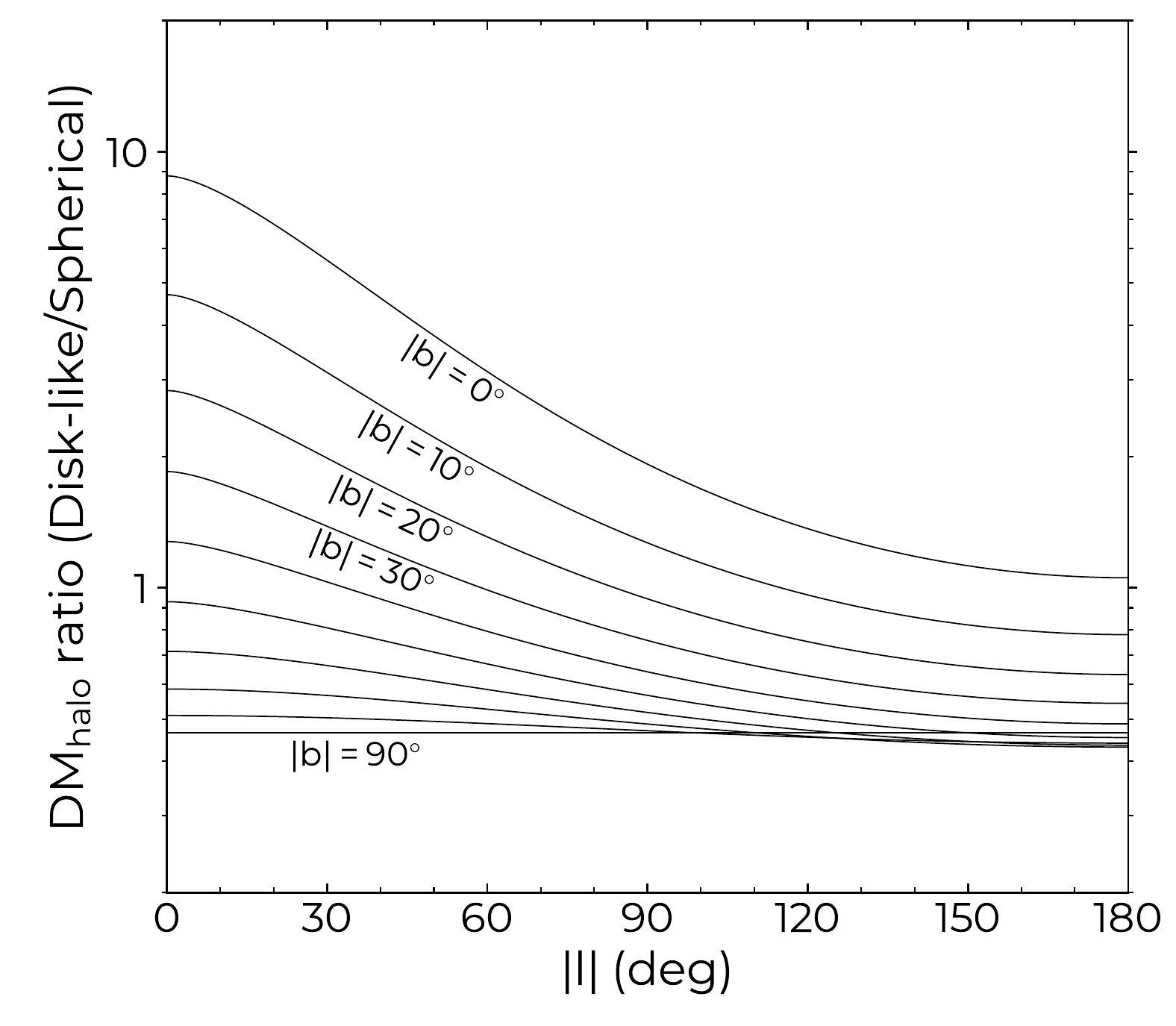}
\caption{Same as Figure \ref{fig:DM_halo}, but the DM ratio of the
  disk-like halo to the spherical component is shown.}
\label{fig:DM_halo_ratio}
\end{figure}

Based on our new hot gas halo model, here we aim to provide a
convenient analytic formula of the hot gas halo DM to any given
direction to an extragalactic object.  We calculated a full-sky map
for ${\rm DM}_{\rm halo}$ by integrating $n_e$ until the sightline
intersects the sphere of the virial radius $r=r_{\rm vir}$.  Figure
\ref{fig:DM_halo} describes the derived DM profile for selected
Galactic latitudes.  
The halo DM of our model spans the range
${\rm DM}_{\rm halo}=30$--$245\:\, {\rm pc\:cm^{-3}}$ over the whole sky, with a mean of
$43\:\, {\rm pc\:cm^{-3}}$.  Figure
\ref{fig:DM_halo_ratio} shows the ratio of the two components of the
halo DM, disk-like to spherical, and it ranges in $0.4$--$9$ over the
full-sky region, which demonstrates the highly non-spherical nature of
our model.

The fitting formula of the halo DM as a function of the Galactic
coordinate is obtained with the $7$th-order polynomial
as
\begin{eqnarray}
\label{eq:polynomial}
{\rm DM_{\rm halo}}=\sum_{i,\,j=0}^{n}c_{ij}\,|l|^i\,|b|^j \ ,
\end{eqnarray}
where $c_{ij}$ is the fitting coefficient in units of ${\rm
  pc\:cm^{-3}}$, $l$ and $b$ are Galactic coordinates measured in
radians, and $n=7$. The fitting result is summarized in Table
\ref{tab:coefficients}. We confirm that this formula reproduces the
theoretical prediction within $4$\% accuracy, and the regions of an
accuracy better than 1\% amount to 98\% of the entire sky.  This
formula for ${\rm DM}_{\rm halo}$ can be used in combination with
existing ${\rm DM}_{\rm ISM}$ models (\citetalias{Cordes2002} and
\citetalias{Yao2017}) to estimate the total DM by electrons in the MW.
In order to separate the ${\rm DM}_{\rm halo}$ contributions by disk-like and spherical component, we also show a fitting result only for the spherical halo component with $n=3$ in Table
\ref{tab:coefficients}, which achieves a higher model accuracy (within $2$\%) due to a smaller directional dependence of the spherical halo.

\section{Discussion}
\label{sec:Discussion}

\subsection{Relation to the Warm Electron Models}
\label{subsec:Discussion ISM}

\begin{figure}[t]
\epsscale{1.15}
\plotone{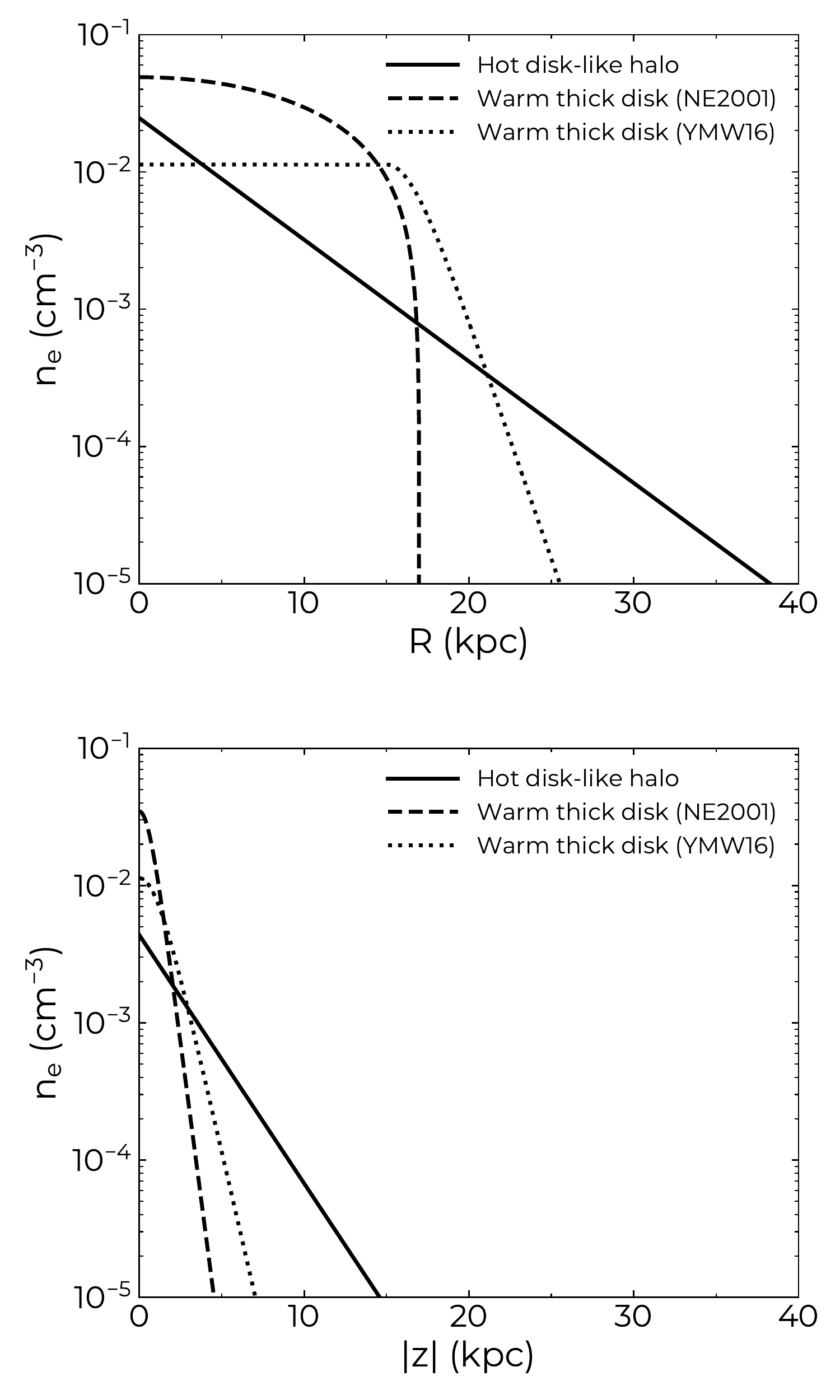}
\caption{{\it Top panel}: electron density plotted against
  Galactocentric radius in the in-plane ($R$-axis) direction from the
  Galactic center for the warm thick disk models and the hot disk-like
  halo; {\it bottom panel}: electron density plotted against distance
  in the vertical ($z$-axis) direction from the Sun.}
\label{fig:disk models}
\end{figure}

The spatial distributions of WIM in the Galactic ISM have been modeled
by the observed DMs toward Galactic radio pulsars
\citep{Taylor1993,Cordes2002,Cordes2003,Yao2017}, and the diffuse
component of WIM is known to distribute in the so-called ``thick disk"
with a vertical scale height of $\lesssim2$ kpc and a mid-plane
electron number density of $\sim0.01\:\rm cm^{-3}$ (see, e.g.,
\citealt{Readhead1975,Reynolds1989,Gaensler2008,Savage2009}).  Figure
\ref{fig:disk models} illustrates the density profile of warm thick
disks (\citetalias{Cordes2002} and \citetalias{Yao2017}) in comparison
with our hot disk-like halo.  Since the gas distribution of the hot
disk-like halo component evidently overlaps with those of thick disk
models, there is a possibility that the hot disk-like halo has already
been taken into account partly in the modeling of the thick disk by
\citetalias{Cordes2002} and \citetalias{Yao2017}.

\begin{figure}[]
\epsscale{1.15}
\plotone{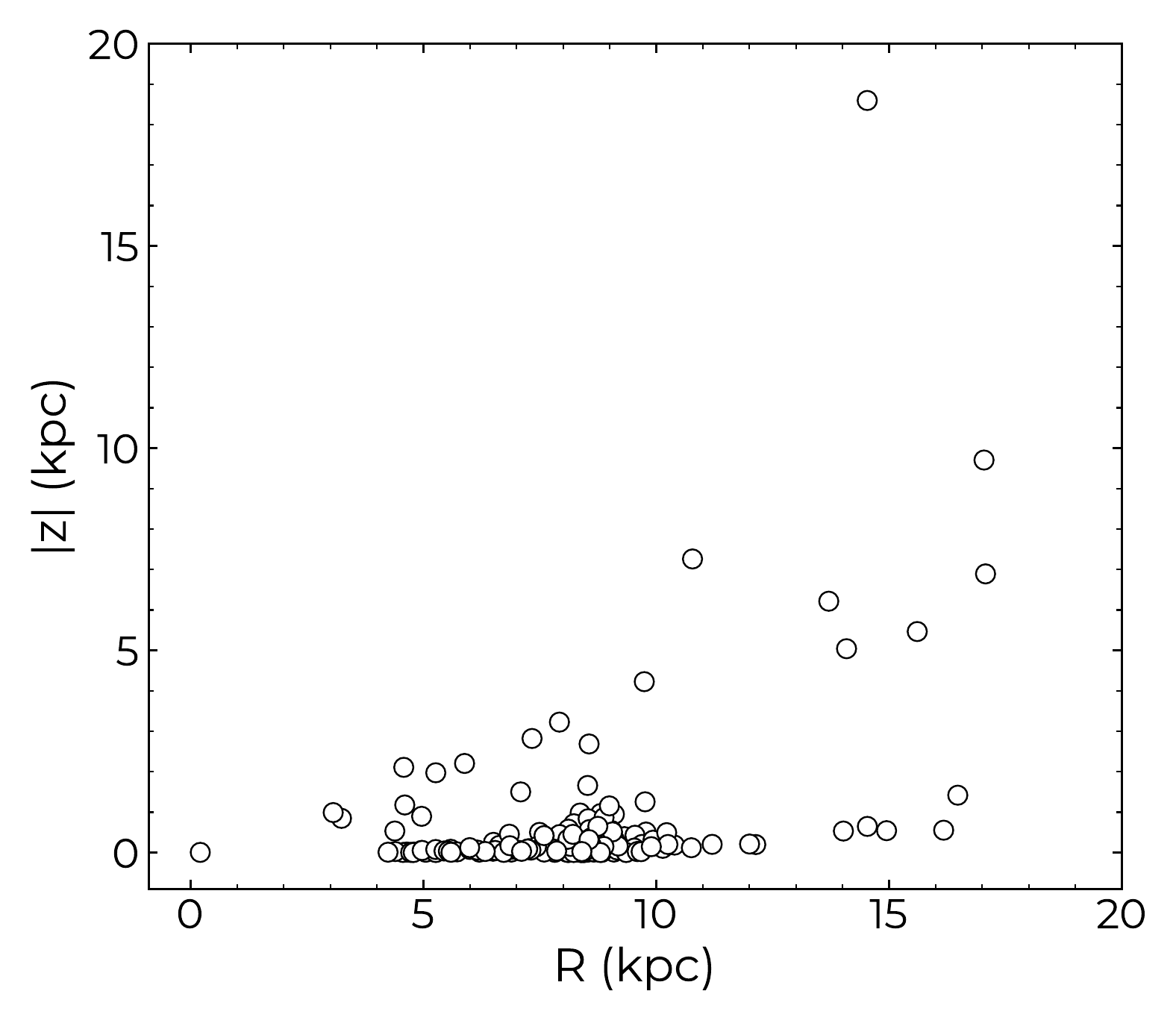}
\caption{Spatial distribution of the $189$ Galactic pulsars with
  DM-independent distances used to constrain the warm thick disk model
  of \citetalias{Yao2017}, which include a smaller ($N=112$) sample of
  pulsars used for \citetalias{Cordes2002}. The direction and distance
  information are adopted from Tables A1--A5 of \citetalias{Yao2017}.}
\label{fig:psr_dist}
\end{figure}

\begin{figure}[]
\epsscale{1.15}
\plotone{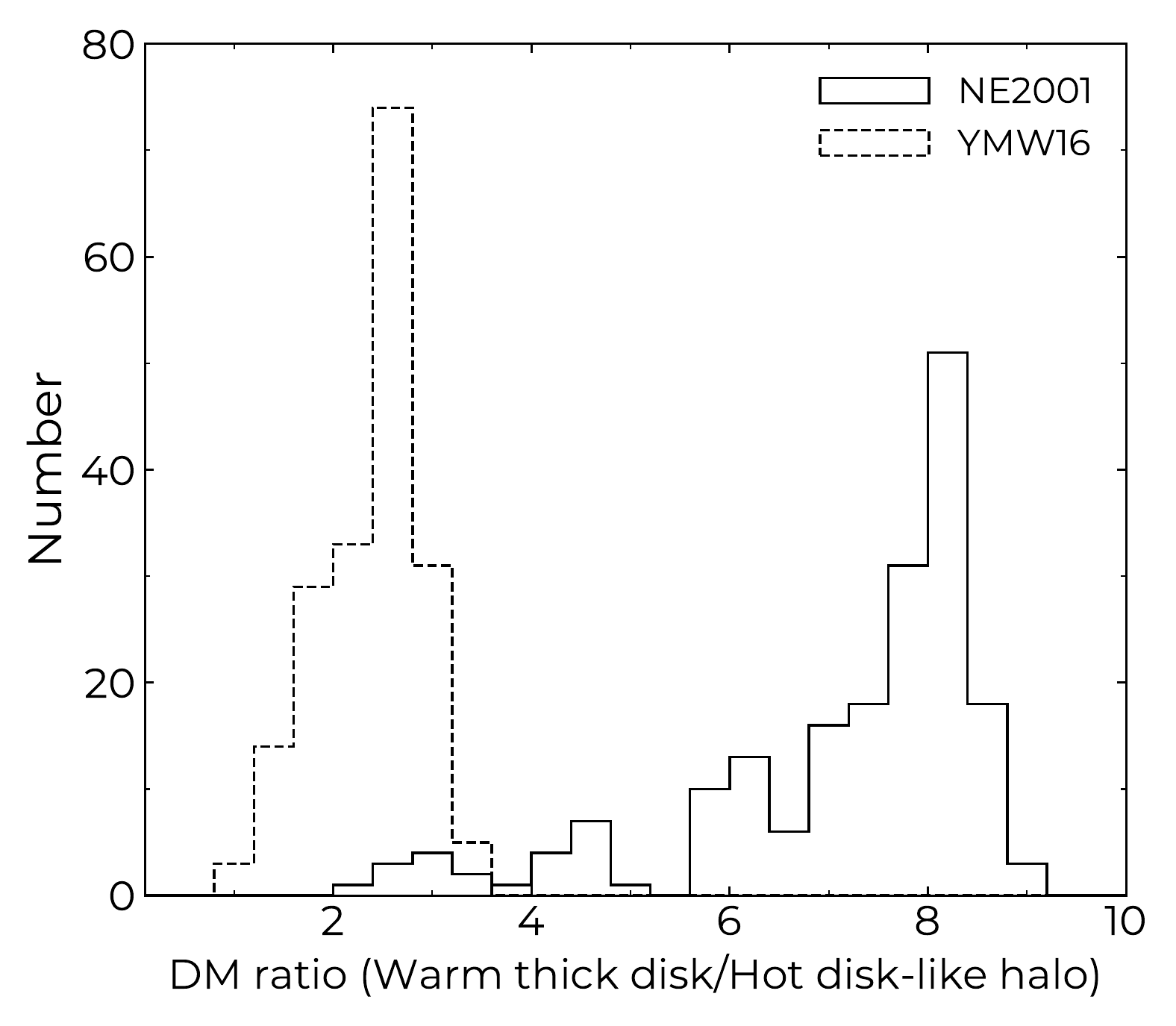}
\caption{Histograms of DM ratio of the warm thick disk to the hot
  disk-like halo for the full sample of $189$ Galactic pulsars shown
  in Figure \ref{fig:psr_dist}.}
\label{fig:psr_DM}
\end{figure}

In Figure \ref{fig:psr_dist} we show the spatial distribution of 189
Galactic pulsars having independent distance constraints (mostly by
parallax measurements) that have been used to model the warm thick
disks (\citetalias{Yao2017}, see also the ATNF Pulsar Catalogue documented in \citealt{Manchester2005}).
The average distance from the Sun to those pulsars is $3.4$ kpc, and
most of them lie in the vicinity of the Galactic plane ($|z|\lesssim2$
kpc and $5$ kpc $\lesssim R\lesssim15$ kpc).  Figure \ref{fig:psr_DM}
compares the predicted DM for these pulsars by the warm thick disks
and the hot disk-like halo.  It clearly indicates that for the
majority of these pulsars, DM contribution from warm thick disks is at
least a few times larger than that from the hot disk-like halo. The
geometrical shapes of the thick disks and our disk-like halo are
significantly different, and it is unlikely that the disk-like halo is
properly taken into account in the warm disk models.  Therefore, we
recommend to simply add ${\rm DM}_{\rm halo}$ of our model (presented
in Section \ref{sec:Analytic formula}) to ${\rm DM}_{\rm ISM}$ of
\citetalias{Cordes2002} or \citetalias{Yao2017} to estimate the total
DM of the MW.

\subsection{Model Uncertainties}
\label{subsec:Model Uncertainties}
Based on our fiducial hot gas halo model, we estimate the mean halo DM of ${\rm DM}_{\rm halo}=43\:\, {\rm pc\:cm^{-3}}$.
We note that this number should
only be considered as a benchmark due to the following systematic uncertainties. First, since the observed EMs can be almost fitted with the disk-like halo component alone, the major source of uncertainty originates from the the modelling of spherical halo component (i.e., the total gas mass of the spherical halo within the virial radius $M_b$). We find that the fraction of the cosmic baryons in the Galactic halo $f_b$, defined as $M_b/M_{\rm vir}=f_b(\Omega_b/\Omega_m)$, needs to be less than unity (a fiducial value of $f_b=0.75$) in order to be consistent with the DM toward the LMC  with the best-fit disk-like halo component being unchanged. If we consider the possible range of $f_b\in[0,1]$ (although $f_b = 0$ is a rather extreme limit), the mean halo DM over the whole sky ranges ${\rm DM}_{\rm halo}=21$--$50\:\, {\rm pc\:cm^{-3}}$. This means that a choice of different $f_b$ (or $M_b$) change the estimate of $|{\rm DM}_{\rm halo}|$ by at most $\lesssim 20\:\, {\rm pc\:cm^{-3}}$. By contrast, statistical uncertainties in the best-fit parameters of the disk-like halo component is negligible.
Secondly, it is found that a choice of larger integration limits corresponding to $r=1.5\,r_{\rm vir}$--$2.0\,r_{\rm vir}$ increases
the mean ${\rm DM}_{\rm halo}$ value only by $14$\%--$26$\%.
Lastly, if we
attribute the scatter seen in the X-ray EM data ($0.4$ dex) to the
fluctuation of the hot gas density, ${\rm DM}_{\rm halo}$ should have a scatter of $0.2$
dex over the whole sky, as ${\rm DM}\propto n_e$ and ${\rm EM}\propto n_e^2$. 

\subsection{Application to Host-identified FRBs}
\label{subsec:Application}

Here we focus on two non-repeating sources FRB 180924
\citep{Bannister2019} and FRB 190523 \citep{Ravi2019} for which the
redshifts are known.  In order to discuss the DM budget for these
sources, we utilize the ${\rm DM}_{\rm IGM}$-$z$ relation
\citep{Ioka2003,Inoue2004,Deng2014}:
\begin{eqnarray}
\label{eq:DM_IGM}
{\rm DM}_{\rm IGM}(z)=\Xi_{\rm IGM}\int_0^z\frac{f_e(z^{\prime})(1+z^{\prime})dz^{\prime}}{\sqrt{\Omega_{m}(1+z^{\prime})^3+\Omega_{\Lambda}}},\nonumber\\
\Xi_{\rm IGM}\equiv\frac{3cH_0\Omega_b f_{\rm IGM}}{8\pi G m_p}\approx 1100\,f_{\rm IGM}\:\, {\rm pc\:cm^{-3}}.
\end{eqnarray}
Here $f_e=1/\mu_e$ is the ionization factor and we neglect the
redshift dependence, and $f_{\rm IGM}$ denotes the fraction of baryons
that reside in the ionized IGM, which has yet to be constrained
well. The current cosmic baryon census
suggests that $f_{\rm IGM}\gtrsim0.6$ \citep{Shull2012} and $f_{\rm
  IGM}$ could be as high as $\sim0.9$ (e.g., \citealt{Fukugita2004})
provided that all the missing baryons ($\sim30$\%) exist as a form of
diffuse IGM. Here we set $f_{\rm IGM}\in[0.6, 0.9]$ as a plausible
range. The systematic errors in our halo model is conservatively taken to be $\pm 20\:\, {\rm pc\:cm^{-3}}$ (see Section \ref{subsec:Model Uncertainties}).

{\it FRB 180924}.---The host is an massive galaxy with stellar mass of
$M_{\ast}\sim2.2\times10^{10}\:M_{\odot}$ at $z\sim0.32$
\citep{Bannister2019}.  The total DM is reported to be ${\rm DM_{obs}}
= 361\: {\rm pc\:cm^{-3}}$ \citep{Bannister2019}, and an upper limit
on DM$_{\rm IGM}$ is obtained by DM$_{\rm IGM} \leq$ DM$_{\rm obs} - $
DM$_{\rm ISM} - $ DM$_{\rm halo}$, where the equality holds when
DM$_{\rm host} = 0$.  The ISM contribution to this direction
[$(l,b)=(0.74^{\circ},-49^{\circ})$] is estimated as ${\rm DM_{ISM}}
= 41$ (\citetalias{Cordes2002}) or $28\: {\rm pc\:cm^{-3}}$
(\citetalias{Yao2017}) by the two different models.  The MW halo
contribution to this direction by our model is ${\rm DM_{halo}}=46_{-20}^{+20}\:
{\rm pc\:cm^{-3}}$, compared to $50$--$80\:
{\rm pc\:cm^{-3}}$ estimated by \citetalias{Prochaska2019}. Compared
with the theoretical value of ${\rm DM}_{\rm IGM} = 320\,f_{\rm IGM}\:
{\rm pc\:cm^{-3}}$ from Eq. (\ref{eq:DM_IGM}), the observation gives a
constraint on $f_{\rm IGM}$. Our DM$_{\rm halo}$ model predicts a
lower value than \citetalias{Prochaska2019}, and hence a weaker constraint of $f_{\rm IGM} <
0.79$--$0.96$ depending on the DM$_{\rm ISM}$ models, which should be
compared with $f_{\rm IGM} < 0.75$--$0.79$ when the high end value of \citetalias{Prochaska2019} is adopted.

{\it FRB 190523}.---The host is a massive
($M_{\ast}\sim5.0\times10^{11}\:M_{\odot}$) galaxy at $z\sim0.66$
\citep{Ravi2019}. The total DM is reported to be ${\rm DM_{obs}} =
761\: {\rm pc\:cm^{-3}}$ \citep{Bannister2019}, with warm ISM
contribution averaged over two models ${\rm DM_{ISM}} = 37$
(\citetalias{Cordes2002}) and $30\: {\rm pc\:cm^{-3}}$
(\citetalias{Yao2017}) for the FRB direction
$(l,b)=(117^{\circ},44^{\circ})$. The MW halo DM of our model is ${\rm
  DM_{halo}}=32_{-20}^{+20}\: {\rm pc\:cm^{-3}}$. Compared with the theoretical
${\rm DM}_{\rm IGM} = 682\,f_{\rm IGM}\: {\rm pc\:cm^{-3}}$, $f_{\rm
  IGM}$ is constrained to $f_{\rm IGM} <0.99$--$1$ using our halo DM
model, while $f_{\rm IGM} < 0.94$--$0.95$ is derived using the high end
value of \citetalias{Prochaska2019}.

\section{Conclusions}
\label{sec:Conclusions}

In this study, we constructed a new model for DM associated with the
extended hot gas halo in the MW, by taking into account the recent
diffuse X-ray observation. Our hot gas halo model comprises of the two
components: disk-like and spherical halo.  The former is suggested by
the recent diffuse X-ray observations, while the latter is
theoretically introduced to make the total baryonic halo mass
consistent with the cosmic baryon-to-dark-matter ratio.  The radial
profile of the spherical component is modeled by an isothermal gas
under dynamical equilibrium with the dark matter halo potential of the
MW. It is shown that the inclusion of the disk-like component is
essential to explain the directional dependence of the observed EMs,
which is in contrast to the previous models considering only the
spherical halo.

Based on the newly proposed hot gas halo density profile, we derive
the halo DM along any line of sight.  Our model predicts a full range of ${\rm DM}_{\rm halo}=30$--$245\: {\rm pc\:cm^{-3}}$ over the whole sky, with a mean of $43\: {\rm pc\:cm^{-3}}$, which is slightly
higher than the prevailing value ($30\: {\rm pc\:cm^{-3}}$) based on
cosmological simulations \citep{Dolag2015}, but lower than the range
preferred by a recent model of \citetalias{Prochaska2019} ($50$--$80\: {\rm pc\:cm^{-3}}$).
We provide a convenient analytic formula for the MW halo DM,
which enables an easy estimate of ${\rm DM}_{\rm halo}$ along any siteline toward extragalactic sources.

With the advent of large field-of-view surveys, such as CHIME and Apertif \citep{vanLeeuwen2014}, the number of
nearby FRBs with ${\rm DM}_{\rm obs}\lesssim100\: {\rm pc\:cm^{-3}}$
(e.g., FRB 171020 with ${\rm DM}_{\rm obs}=114\: {\rm pc\:cm^{-3}}$
found by ASKAP; \citealt{Shannon2018,Mahony2018}, FRB 110214 with
${\rm DM}_{\rm obs}=169\: {\rm pc\:cm^{-3}}$ by Parkes;
\citealt{Petroff2019} and FRB 181030.J1054+73 with
${\rm DM}_{\rm obs}=104\: {\rm pc\:cm^{-3}}$ by CHIME; \citealt{Andersen2019}) is expected to increase in the foreseeable
future. Since the total DMs of nearby FRBs might be dominated by the
contribution from Galactic electrons,
the estimate of ${\rm DM}_{\rm ISM}$ and ${\rm DM}_{\rm halo}$ would be more important.

\acknowledgments We are grateful to Shinya Nakashima for providing us
detailed information about the {\it Suzaku} data analysis and the anonymous referee for helpful comments that have significantly improved the quality of the manuscript. SY thanks J. Xavier Prochaska for fruitful discussion during his stay in the University of Tokyo. SY also thanks John H. Livingston for carefully reading the manuscript. SY was
supported by the Research Fellowship of the Japan Society for the
Promotion of Science (JSPS) Grant Numbers {\rm JP17J04010}. TT was
supported by JSPS/MEXT KAKENHI Grant Numbers 18K03692 and 17H06362. This
research made use of Astropy, a community-developed core Python
package for Astronomy \citep{astropy:2013, astropy:2018}.

usepackage{natbib}



\end{document}